\documentclass[bibyear]{aa}  

\usepackage[]{natbib}
\bibpunct{(}{)}{;}{a}{}{,}	    

\usepackage{graphicx}
\usepackage{txfonts}
\usepackage[]{hyperref}
\usepackage{mathabx}

\newcommand{\teff}{t_\text{eff}}
\newcommand{\epsilonparall}{\epsilon_\|}
\newcommand{\epsilonperp}{\epsilon_\bot}
\newcommand{\piparall}{\pi_\|}
\newcommand{\piperp}{\pi_\bot}
\newcommand{\sigmapi}{\sigma_{\pi_E}}
\newcommand{\sigmapimin}{\sigma_{\pi_E, \, \text{min}}}
\newcommand{\sigmapimax}{\sigma_{\pi_E, \, \text{max}}}
\newcommand{\sigmaparall}{\sigma_{\piparall}}
\newcommand{\sigmaperp}{\sigma_{\piperp}}
\newcommand{\sigmaextrema}{\sigma_{\pi_E, \pm}}
\newcommand{\tEstar}{t_E^{\star}}
\newcommand{\mjup}{M_{\text{Jup}}}

\begin{document} 

\title{Microlensing planet detection\\via geosynchronous and low Earth orbit satellites}

\author{F. Mogavero \and J. P. Beaulieu}

\institute{Sorbonne Universités, UPMC Paris 6 et CNRS, UMR 7095,
Institut d’Astrophysique de Paris, 98 bis bd Arago, 75014 Paris, France\\
\email{mogavero@iap.fr,beaulieu@iap.fr}}

\date{Received \today; accepted }
    
\abstract{Planet detection through microlensing is usually limited by a well-known 
   degeneracy in the Einstein timescale $t_E$, which prevents mass
   and distance of the lens to be univocally determined. It
   has been shown that a satellite in geosynchronous orbit could provide
   masses and distances for most standard planetary events ($t_E \approx 20$ days)
   via a microlens parallax measurement. This paper extends the analysis to
   shorter Einstein timescales, $t_E \approx 1$ day, when dealing with
   the case of Jupiter-mass lenses. We then study the
   capabilities of a low Earth orbit satellite on even shorter timescales, $t_E \approx
   0.1$ days. A Fisher matrix analysis is employed to predict how the 1-$\sigma$
   error on parallax depends on $t_E$ and the peak magnification
   of the microlensing event. It is shown that a geosynchronous satellite could
   detect parallaxes for Jupiter-mass free floaters and discover planetary
   systems around very low-mass brown dwarfs. Moreover, a low Earth orbit satellite
   could lead to the discovery of Earth-mass free-floating planets. Limitations
   to these results can be the strong requirements on the photometry, the effects of
   blending, and in the case of the low orbit, the Earth's umbra.}


\maketitle
%

\section{Introduction}

   The fundamental quantity that is routinely measured in a microlensing event
   is the Einstein timescale $t_E$,
   \begin{equation}
   \label{tE}
   t_E = 1.1 \, \text{days} \, {\left( \frac{M}{\mjup} \right)}^{1/2}
   {\left( \frac{x(1-x)}{0.25} \right)}^{1/2}
   {\left( \frac{D_S}{8 \, \text{kpc}} \right)}^{1/2}
   \left( \frac{200 \, \text{km/s}}{V} \right)
   \end{equation}
   where $x \equiv D_L/D_S$. It depends on the lens mass $M$, the lens and source
   distances from Earth $D_L$ and $D_S$, and the lens-source relative
   velocity $V$. Consequently, the lens mass cannot be inferred without any knowledge
   of $x$ and $V$.
   This degeneracy in $t_E$ can be resolved via the equation $M =
   \theta_E/\kappa \pi_E$, with $\kappa$ a constant, if the angular Einstein
   radius $\theta_E = t_E V/D_L$ and the microlens parallax
   $\pi_E = 1 \, \text{AU} (D_L^{-1} - D_S^{-1})/\theta_E$,
   \begin{equation}
   \label{piE}
   \pi_E = 4.3 \left( \frac{1 \, \text{day}}{t_E} \right) \left( \frac{1-x}{0.5} \right)
   \left( \frac{200 \, \text{km/s}}{V} \right)
   \end{equation}
   can both be measured from the event light curve (\citealt{Gould2013}, from now
   on G13).
   The quantity $\theta_E$ is measured via finite source effects
   in high-magnification single-lens events and in a large number of the planetary ones. 
   G13 has shown that for planetary events
   ($t_E \approx 20$ days) with peak
   amplifications $A_{\text{max}} \gtrsim 20$, a satellite
   placed in geosynchronous orbit would measure the microlens parallax $\pi_E$,
   providing lens masses. This result is especially important in the perspective of the
   Wide Field Infrared Space Telescope (WFIRST) mission \citep{Barry2011}: a
   geosynchronous orbit is currently a strong alternative to the second Lagrangian
   point (L2) \citep{Spergel2015}.
   In particular, G13 studies the regime where the satellite orbital period $P$ is much 
   shorter than the shortest timescale of the microlens event, $\teff \equiv \beta t_E$,
   with $\beta$ the lens-source impact parameter in units of $\theta_E$. The following
   scaling law for the relative error on $\pi_E$ is predicted:
   \begin{equation}
   \label{gould}
   \frac{\sigmapi}{\pi_E} \, \propto \, {t_E}^{1/2} \, \beta \, R^{-1}, \quad
   P \ll \beta t_E \, \, \text{and} \, \, \beta \ll 1
   \end{equation}
   where $R$ is the satellite orbital radius.
   The sensitivity to $\pi_E$ increases towards shorter Einstein
   timescales. However, Eq. \eqref{gould} does not apply when $P \gtrsim \beta t_E$,
   as correlations between $\pi_E$ and the other model parameters, such as
   $\beta$ and $t_E$, start to affect the error on $\pi_E$.\\
   In this paper, after briefly setting up the Fisher matrix analysis in Sect. \ref{Fisher_analysis},
   we extend the analysis of G13 to shorter Einstein timescales in
   Sect. \ref{geosynchronous}, to cover events raised by Jupiter-mass
   lenses ($t_E \approx 1$ day).
   Then, in Sect. \ref{LEO} we study the capabilities of a low Earth
   orbit satellite to detect microlens parallax in case of Earth-mass lenses ($t_E
   \approx 0.1$ days).

\section{Fisher matrix analysis}
\label{Fisher_analysis}

   The Fisher information matrix provides a way to predict, in a Bayesian
   framework, constraints on the best fit of a theoretical model to observational
   data. It allows the covariance matrix of the model parameters to be estimated
   via their posterior probability distribution \citep{Sellentin2014}. The Fisher
   matrix is a fast, analytical alternative to time-consuming methods such as
   the Markov-Chain Monte Carlo, even though its reliability is not always
   guaranteed \citep{Vallisneri2008}.
   
   The physical observable of a microlensing event is the light flux $F$,
   \begin{equation}
   \label{flux}
   F = F_s \, A + F_b = \overline{F} \, [(1- \nu) \, A + \nu]
    \end{equation}
   where $F_s$, $F_b$, and $\overline{F} \equiv F_s + F_b$ are the source, blending,
   and baseline fluxes, and $\nu \equiv F_b/(F_s + F_b)$ defines the
   blending ratio. In a point source-point lens (PSPL) model, the amplification factor $A$
   is given by $A(u) = (u^2 + 2)/u \sqrt{u^2 + 4}$,
    where $u$ is the magnitude of the lens-source separation vector $\vec{u}$ as seen
    by the observer. For an inertial observer, $u = \sqrt{\tau^2 + \beta^2}$,
    where $\tau \equiv (t-t_0)/t_E$ and $t_0$ is the peak time.
    The inertial observer model is thus described by five parameters: $\vec{\theta} =
    (\overline{F}, \nu, t_0, t_E, \beta)$.
    
    Under the assumption of Gaussian errors,
    the Fisher matrix elements are given by \citep{Sellentin2014}
   \begin{equation}
   \label{FisherMatrix}
   \mathcal{F}_{ij} = \sum_{k = 1}^N \frac{1}{\sigma_k^2}
   \frac{\partial F_k}{\partial \theta_i} \frac{\partial F_k}{\partial \theta_j}, \quad
   \sigma^2 = \sigma_0^2 \, [(1- \nu) \, A + \nu]
    \end{equation}
    where the index $k$ spans the set of $N$ independent observations, and
    $\sigma$ is the flux error. We have assumed Poisson noise-limited photometry
    with $\sigma_0 \simeq (\ln 10/2.5) \overline{F} \sigma_m$, where
    $\sigma_m$ is the magnitude error at the light curve baseline.
    In the case of continuous observations,
    i.e. $f \beta t_E \gg 1$, where $f$ is the number of observations per unit time,
    the sum appearing in equation \eqref{FisherMatrix} can be approximate by an
    integral over time,
    \begin{equation}
    \label{sum_to_integral}
    \sum_{k = 1}^N \longrightarrow f t_E \int_{- \infty}^{+ \infty} d\tau.
    \end{equation}
    Since the parallax signal contributing to the matrix $\mathcal{F}$ comes
    from a few $\teff$ near $\tau = 0$, the boundaries of integration can be safely
    extended to infinity, as shown in Eq. \eqref{sum_to_integral}, if the observations
    last more than a few event timescales\footnote{Substitution
    \eqref{sum_to_integral} causes the element $\mathcal{F}_{\overline{F}
    \, \overline{F}}$ to diverge, meaning that an arbitrary
    precision on $\overline{F}$ can be attained by observing enough in
    the baseline of the light curve. The parameter $\overline{F}$ can thus be safely
    considered as a constant and can be removed from the Fisher analysis.}.
    Once the Fisher matrix is computed, the covariance matrix of the model parameters
    is given by its inverse, $\mathcal{F}^{-1}$ \citep{Sellentin2014}.

    \subsection{Parallax effect for an Earth orbit satellite}
    Observations from an Earth orbit satellite are affected by
    the parallax effect of its non-inertial motion,
    in the same way as for a telescope
    on Earth \citep{Hardy1995}. We thus consider an observatory
    in circular orbit around Earth with radius $R$ and period $P$ ($\omega =
    2 \pi/P$ as its angular velocity). If $\lambda$ denotes the latitude of the
    source star with respect to the plane of this orbit, the projection
    of the satellite trajectory onto the plane of the sky is an ellipse with
    semimajor and semiminor axes $R$ and $R \sin(\lambda)$, respectively.
    Following G13, one then defines $\epsilonparall = \epsilon \equiv
    R/1 \, \text{AU}$ and $\epsilonperp \equiv R \sin(\lambda)/1 \, \text{AU}$.
    Let $\theta$ denote the direction of the lens-source relative
    motion in the plane of the sky with respect to the projected major axis of
    the satellite orbit. The lens-source separation vector seen by the satellite
    is given by $\vec{u} = (\tau \cos\theta - \beta \sin\theta + \epsilonparall
    \pi_E \cos(\omega t_E \tau + \varphi), \tau \sin\theta + \beta \cos\theta +
    \epsilonperp \pi_E \sin(\omega t_E \tau + \varphi))$, where $\varphi$ is the satellite
    orbital phase with respect to the peak time $t_0$ (\citealt{Hardy1995},
    G13). If one introduces the microlens parallax vector $\vec{\pi_E} = (\piparall,
    \piperp) = \pi_E (\cos\theta, \sin\theta)$, the non-inertial observer model is
    described by seven parameters: $\vec{\theta} = (\overline{F},
    \nu, t_0, t_E, \beta, \piparall, \piperp)$. The two additional flux derivatives
    $\partial F/\partial \piparall$ and $\partial F/\partial \piperp$ appearing in the
    Fisher matrix \eqref{FisherMatrix} are proportional to $\partial u/\partial \piparall$
    and $\partial u/\partial \piperp$, respectively, with
    \begin{equation}
    \begin{split}
    \label{parallax}
    &u \frac{\partial u}{\partial \piparall} \simeq \epsilonparall \tau \cos(\omega t_E
    \tau + \varphi) + \epsilonperp \beta \sin(\omega t_E \tau + \varphi), \\
    &u \frac{\partial u}{\partial \piperp} \simeq - \epsilonparall \beta \cos(\omega t_E
    \tau + \varphi) + \epsilonperp \tau \sin(\omega t_E \tau + \varphi).
    \end{split}
    \end{equation}
    The approximate equalities mean we neglect terms that are linear in $\pi_E$.
    Moreover,  one can assume $u \simeq \sqrt{\tau^2 + \beta^2}$. These approximations
    are justified as long as $\epsilon \, \pi_E \, \beta^{-1} \ll 1$. In the following
    analysis, the above inequality is safely verified. G13 also assumes its validity.

   \subsection{Sensitivity to $\pi_E$}
    Once the Gaussian approximation to the posterior probability distribution of
    $\piparall$ and $\piperp$ is known via the Fisher matrix, one can forecast the
   sensitivity of microlensing observations to $\pi_E$ through standard error
    propagation. Since small variations in the parallax parameters are related by
    $\delta \pi_E \simeq \cos\theta \, \delta \piparall + \sin\theta \, \delta \piperp$,
    the 1-$\sigma$ error on $\pi_E$ is given by the equation
    \begin{equation}
    \label{sigmapi}
    \sigma_{\pi_E}^2 = \cos^2(\theta) \, \sigmaparall^2 + \sin^2(\theta) \,
    \sigmaperp^2 + \sin(2\theta) \, \text{cov}(\piparall, \piperp),
    \end{equation}
    where $\text{cov}(\piparall, \piperp)$ is the covariance between the two
    parallax parameters. In principle, $\sigmapi^2$ depends on
    both $\theta$ and $\varphi$. 
    
    To present results that are independent of the geometry of the event,
    we summarise the information contained into Eq. \eqref{sigmapi} by
    finding the extrema of $\sigmapi^2(\theta, \varphi)$
    over $\theta, \varphi \in [0, 2\pi]$. To do this, one first notices
    that, according to Eq. \eqref{parallax}, $\sigmaparall^2$, $\sigmaperp^2$, and
    $\text{cov}(\piparall, \piperp)$ do not depend on $\theta$, but only on
    $\varphi$. One can thus analytically find the extrema of
    $\sigmapi^2(\theta, \varphi)$ over the $\theta$ range,
    \begin{equation}
    \label{extrema}
    \sigmaextrema^{2}(\varphi)
    = \frac{\sigmaparall^2 + \sigmaperp^2}{2} \pm \frac{\sqrt{(\sigmaparall^2
    - \sigmaperp^2)^2 + 4 \, \text{cov}(\piparall, \piperp)^2}}{2}
    \end{equation}
    where the plus and minus signs stand for the maxima and minima.
    The values of $\theta$ that correspond to these extrema are given by
    $\tan(2\theta) = 2 \, \text{cov}(\piparall, \piperp)/(\sigmaparall^2 -
    \sigmaperp^2)$ \footnote{The extrema $\sigmaextrema(\varphi)$
    correspond to a lens-source relative motion aligned
    with one of the principal axes of the bivariate Gaussian distribution
    which approximates the posterior probability distribution of $\piparall$
    and $\piperp$ in the Fisher matrix analysis, when one marginalises over the
    remaining parameters.}.
    Then, one can find the extrema of Eq. \eqref{extrema} over
    the $\varphi$ range numerically, and they only depend on
    $t_E$ and $\beta$:
    \begin{equation}
    \sigmapimax^2 \equiv \max_{\varphi \in [0, 2 \pi]}
    \sigma_{\pi_E, \, +}^2(\varphi), \quad \sigmapimin^2 \equiv
    \min_{\varphi \in [0, 2 \pi]} \sigma_{\pi_E, \, -}^2(\varphi).
    \end{equation}
    We note that for $P \ll \beta t_E$, the covariance between
    $\piparall$ and $\piperp$ vanishes, $\sigmaparall^2$
    and $\sigmaperp^2$ become independent of $\varphi$,
    and the extrema of $\sigmapi^2(\theta, \varphi)$
    are simply given by $\max(\sigmaparall^2, \sigmaperp^2)$ and
    $\min(\sigmaparall^2, \sigmaperp^2)$.
    

\section{Geosynchronous orbit satellite}
\label{geosynchronous}

    \begin{figure}
    \centering
    \includegraphics[width=\hsize]{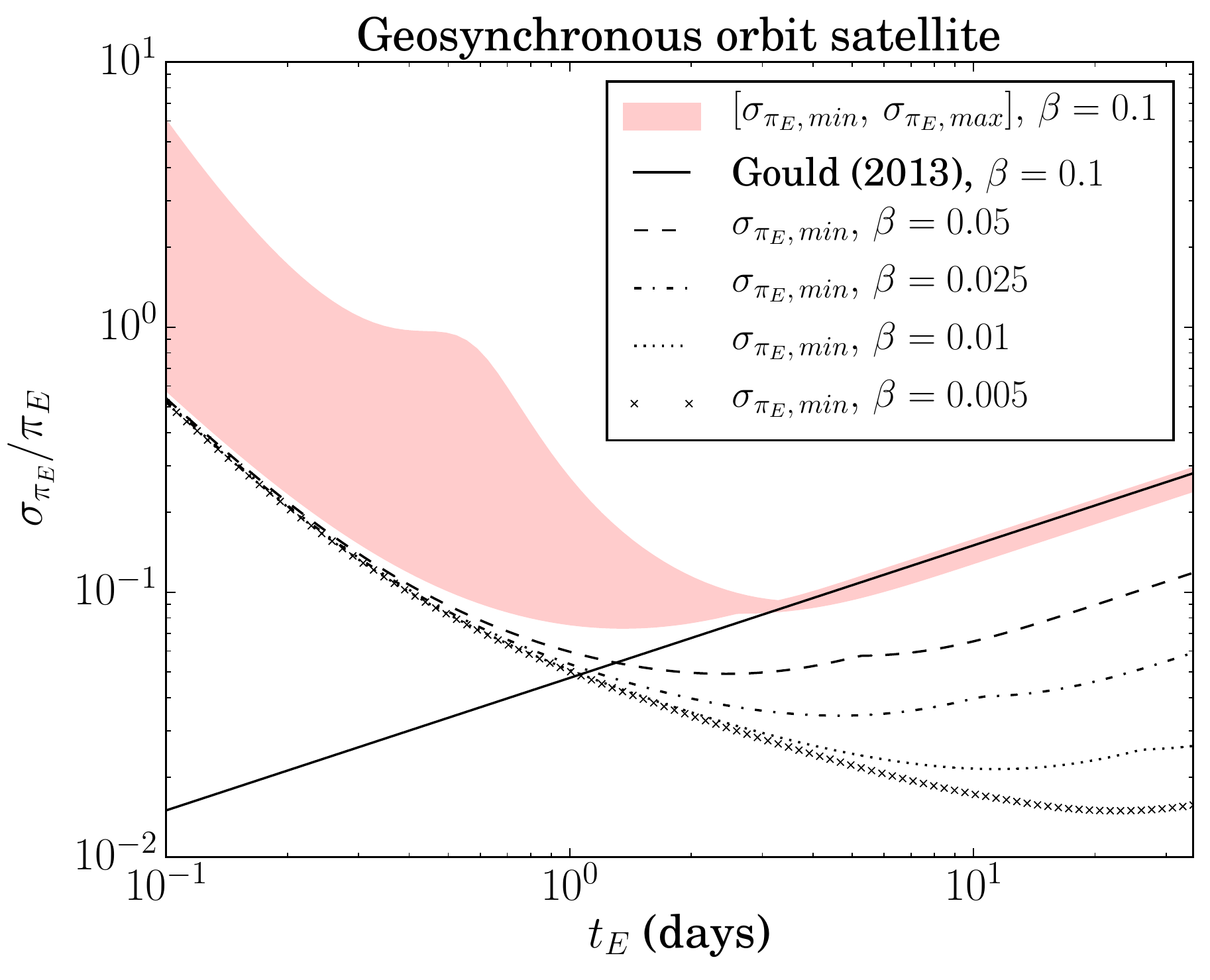}
    \caption{Relative error on $\pi_E$ plotted against $t_E$ for a GSO satellite.
    We assume 3 min exposures, $\sigma_m = 0.01$,
    and zero blending. For $\beta = 0.1$ the region $\sigmapi/\pi_E
    \in [\sigmapimin, \, \sigmapimax]/\pi_E$ is shown, along with the
    G13 prediction. For $\beta = 0.05, 0.025, 0.01, \,
    \text{and} \, 0.005,$ the minimum relative error $\sigmapimin/\pi_E$ is plotted.}
    \label{figure1}
    \end{figure}

    We then consider the case of a geosynchronous orbit (GSO) satellite, orbiting in the
    equatorial plane and targeting the Galaxy bulge, $P = \text{23 h 56 min 4 s}$,
    $R = 6.6 \, R_{\Earth}$, $\lambda = 30^\degree$. We assume $f = 1 \,
    \text{observation}/ 3 \, \text{minutes}$, which is reasonable for a WFIRST-like
    satellite (G13) and a magnitude error at baseline $\sigma_m = 0.01$.
    Figure \ref{figure1} shows the predicted relative error $\sigmapi/\pi_E$
    as a function of the Einstein timescale $t_E$, for zero blending ($\nu = 0$).
    Since, from Eq. \eqref{piE}, $\pi_E$ depends on $D_L$, $D_S$, and $V$,
    we consider a typical disk lens at $D_L = 4 \, \text{kpc}$ and
    $V = 200 \, \text{km/s}$ ($D_S = 8 \, \text{kpc}$ for a source in
    the bulge of the Galaxy).

    For an impact parameter $\beta = 0.1$, which corresponds to a peak amplification
    $A_\text{max} \approx 10$, we plot the region $\sigmapimin \leq \sigmapi
    \leq \sigmapimax$ to take the variations induced by the parameters
    $\varphi$ and $\theta$ into account. Figure \ref{figure1} clearly
    shows two different regimes. For $t_E \gg 1 \, \text{day}$, the relative
    error on $\pi_E$ scales as ${t_E}^{1/2}$, in agreement with Eq. \eqref{gould},
    the analytic prediction of G13. For $t_E \ll 1 \, \text{day}$,
    the relative error increases by decreasing $t_E$, as a result of the correlations
    between $\piparall, \piperp$, and the other parameters $\nu, t_0, t_E,$ and $\beta$.
    In the intermediate region $1 \, \text{day} \leq t_E \leq 3 \, \text{days}$,
    a maximum sensitivity to $\pi_E$ is attained. In particular,
    at $t_E \approx 3 \, \text{days}$, the error range is very
    narrow around $\sigmapi/\pi_E \approx 9\%$. By decreasing $t_E$,
    $\sigmapimax$ increases steeply, while $\sigmapimin$ still decreases
    to a minimum value of about $7\%$ at $t_E \approx 1 \, \text{day}$.
    This means that one can end up with a good sensitivity to $\pi_E$ even at
    $t_E \approx 1 \, \text{day}$, where the maximum relative error is already a
    steep function of the Einstein timescale.
    
    For smaller impact parameters, $\beta < 0.1$, Fig. \ref{figure1} shows the
    minimum relative error $\sigmapimin/\pi_E$. The region of maximum sensitivity
    to $\pi_E$ clearly moves to higher Einstein timescales. If one considers
    the Einstein timescale $\tEstar$ which minimizes $\sigmapimin/\pi_E$, it
    approximately scales according to
    \begin{equation}
    \label{tEstar}
    \tEstar \, \propto \, P \, \beta^{-1}.
    \end{equation}
    Indeed, it separates the two regimes $\beta t_E \ll P$ and $\beta t_E \gg P$.
    To estimate the minimum relative error for a given $\beta$,
    one can write that, approximately, $\sigmapimin/\pi_E \, \propto \,
    {t_E}^{1/2} \, \beta \, R^{-1} g(\beta t_E \omega)$, where $g(y)$
    is a function that tends to $1$ for $y \rightarrow \infty$, to match
    Eq. \eqref{gould}. Since Eq. \eqref{tEstar} implies $\beta \tEstar
    \omega = \text{const}$, one obtains
    \begin{equation}
    \label{sigmapiE_min}
    \frac{\sigmapimin}{\pi_E} \left( \tEstar \right) \, \propto \, P^{1/2} \,
    \beta^{1/2} \, R^{-1}.
    \end{equation}
    Equations \eqref{tEstar} and \eqref{sigmapiE_min} are in good agreement with
    Fig. \ref{figure1}. The maximum sensitivity to $\pi_E$ increases slowly by
    decreasing the impact parameters $\beta$, while the range of the Einstein
    timescales that allow for a parallax detection clearly widens.

    \begin{figure}
    \centering
    \includegraphics[width=\hsize]{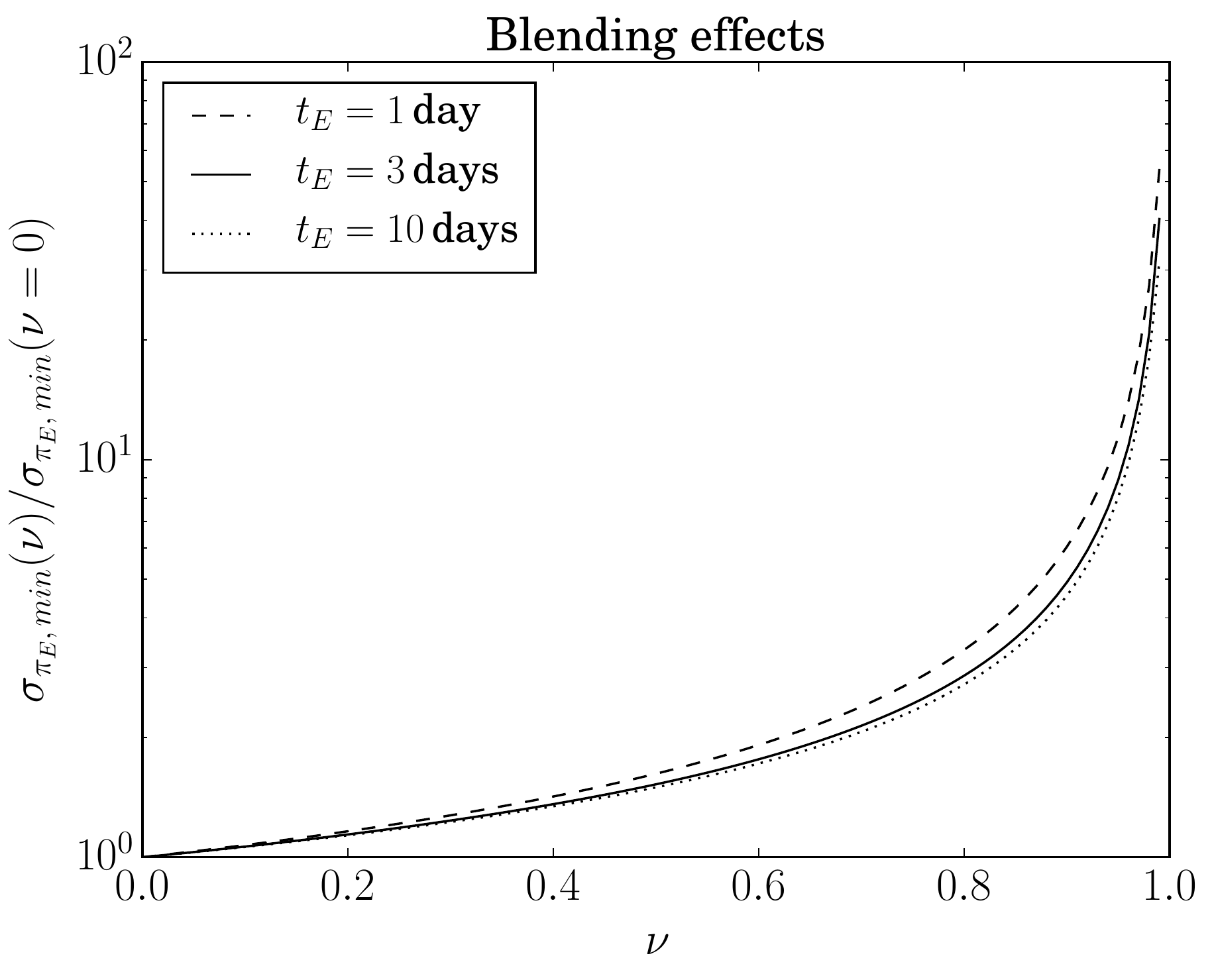}
    \caption{Ratio $\sigmapimin(\nu)/\sigmapimin(\nu=0)$
    against $\nu$ for a GSO satellite and $\beta = 0.1$, with the same
    assumptions as in Fig. \ref{figure1}. Three Einstein timescales are shown: $t_E =$ 
    1, 3, and 10. Clearly the error diverges for $\nu \rightarrow 1$.
    }
    \label{figure3}
    \end{figure}

    From the above analysis and Eq. \eqref{tE}, it follows that a GSO satellite is naturally
    optimized to measure microlens parallax $\pi_E$ in events raised by free-floating
    objects with masses spanning from a fraction to a few dozens $\mjup$.
    The peak amplifications for such a measure can be as low as 10, or even
    somewhat lower for closer lenses. If,
    additionally, the Einstein angle $\theta_E$ can be measured via finite
    source effects (starting at $A_{\text{max}} \approx 50$ for a Jupiter-mass
    lens and a Sun-like source, see Eq. \ref{z}), the lens mass can be inferred for this class
    of events.\\
    Analogously, since for a bulge source the Einstein radius scales as
    $R_E = t_E V = 0.12 \, \text{AU} \, {(M/\mjup)}^{1/2} {(x(1-x)/0.25)}^{1/2}$,
    a GSO satellite is naturally optimized to discover planets in tight orbits
    around low-mass brown dwarfs\footnote{\citealt{Han2013} report the discovery
    of such a system. However, the mass ratio of this binary
    appears too high to envisage that the companion formed in a protoplanetary
    disk around the host.}.
    In particular, as a matter of speculation, it could lead to the discovery
    of miniature planetary systems around planetary-mass brown
    dwarfs\footnote{These could also be interpreted as
    free-floating planet-moon systems \citep{Bennett2014}.}. In fact, disks
    have been found in the past fifteen years to surround brown dwarfs with masses
    in the range $5-15 \, \mjup$ (see, for example, \citealt{Luhman2005feb,
    Luhman2005dec}). In particular, the disk around OTS 44 has
    been estimated at roughly $30 \, M_{\Earth}$ \citep{Joergens2013}.
    The question of whether planets can form out of such disks naturally arises, but
    very little is known about these hypothetical planetary systems, even from a theoretical
    point of view. Gravitational microlensing could thus lead to fundamental discoveries
    in this field.
    
    \subsection{Blending}
    G13 adopts $\sigma^2 = \sigma_0^2 \, A$ in Eq. \eqref{FisherMatrix},
    meaning that it neglects the blending flux $F_b$ by stating that one is
    ``only concerned with the scaling of the errors when the source is relatively
    highly magnified''. Actually, Eq. \eqref{flux} shows that the information
    contained in the Fisher matrix thanks to
    the source flux amplification is scaled by a factor $1-\nu$; i.e., the light
    curve provides less information about the model parameters if the blending factor
    $\nu$ is bigger. This is illustrated in Fig. \ref{figure3}, where we plot the ratio
    $\sigmapimin(\nu)/\sigmapimin(\nu=0)$ against $\nu$ for $\beta = 0.1$ and 
    different values of the Einstein timescale $t_E$. The error roughly doubles at
    $\nu \approx 0.6$, and the detection of $\pi_E$ becomes quite
    hopeless for $\nu \gtrsim 0.9$. Blending can thus highly affect the
    sensitivity to $\pi_E$ and it must always be taken into account in survey planning.
    It is also important to point out that, even for very low values of the blending factor,
    $\nu \rightarrow 0$, this parameter still affects the sensitivity to the other ones.
    Indeed, a small variation $\delta\nu$ produces a change in the total flux $F$
    according to $|\delta F|/F \simeq (A-1)\delta\nu/[(1-\nu)A + \nu]$. For
    $\nu \rightarrow 0$ and $A \gg 1$, one obtains $|\delta F|/F \simeq \delta\nu$,
    which means that the contribution of $\nu$ to the Fisher matrix is different
    from zero even for vanishing blending factors.
    Gould's assumption in G13 is only valid for $\nu \rightarrow 0$ and $P \ll \beta t_E$,
    because in this regime the correlations between $\nu$ and the parallax
    parameters $\piparall$ and $\piperp$ turn out to vanish\footnote{We note
    that, if for a given event $\nu = 0$ holds exactly,
    the parameter $\nu$ must be removed from the corresponding Fisher analysis,
    and the present predictions provide an upper bound to $\sigmapi$.}.


\section{Low Earth orbit satellite}
\label{LEO}

    \begin{figure}
    \centering
    \includegraphics[width=\hsize]{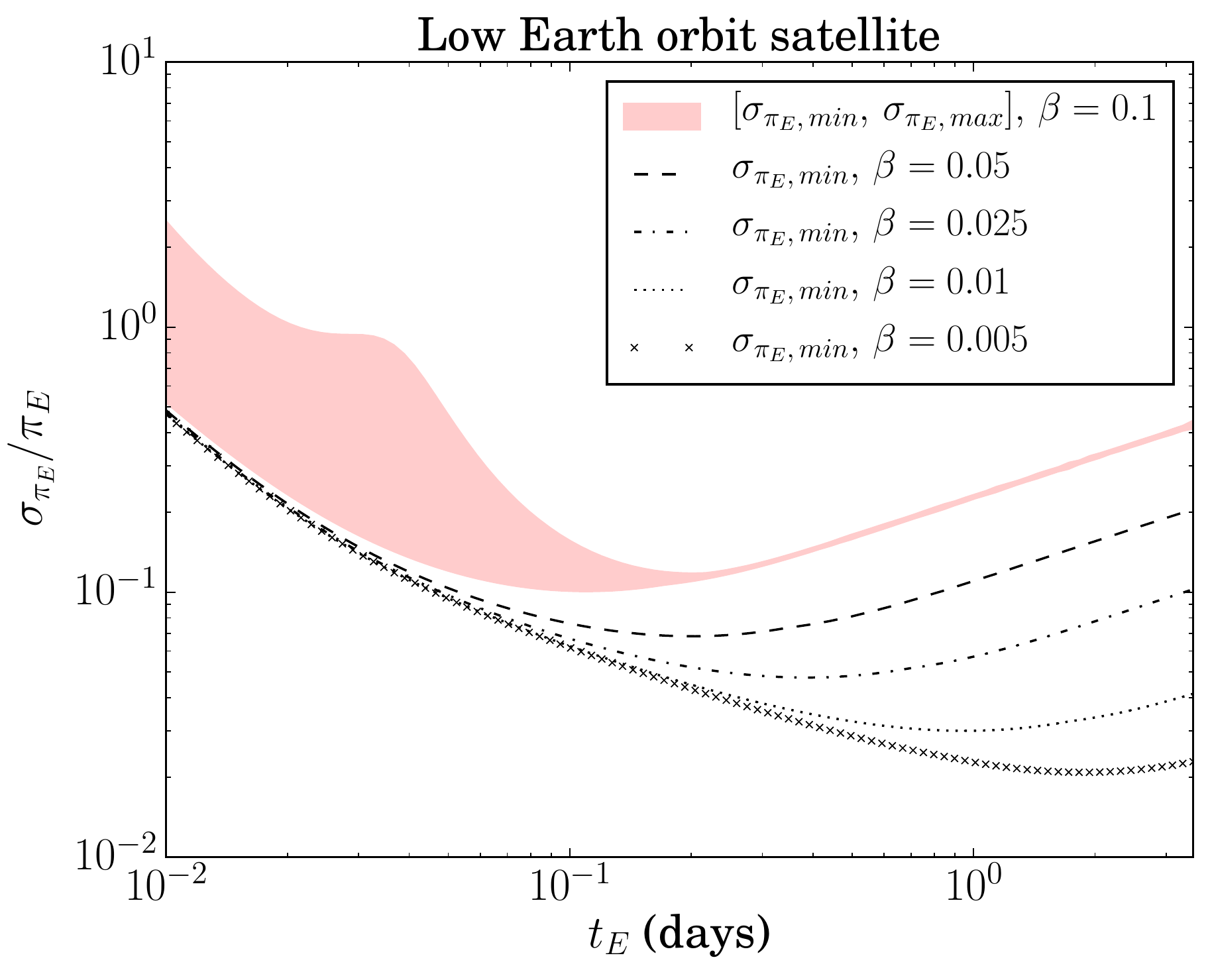}
    \caption{Relative error on $\pi_E$ plotted against $t_E$ for a LEO satellite
    at 550 km above the Earth's surface. We assume 3 min exposures,
    $\sigma_m = 0.01$,
    and zero blending. For $\beta = 0.1$ the region $\sigmapi/\pi_E
    \in [\sigmapimin, \, \sigmapimax]/\pi_E$ is shown. For $\beta = 0.05, 0.025, 0.01, \,
    \text{and} \, 0.005$, the minimum relative error $\sigmapimin/\pi_E$ is plotted.}
    \label{figure4}
    \end{figure}

    Using a space telescope in low Earth orbit (LEO), such as the HST, measuring
    microlens parallaxes was first proposed by \citet{Honma1999}. In G13 Gould
    argues that this would generally be useless, since the corresponding
    $\epsilon = R/1 \, \text{AU}$ is too small. However, substituting
    Kepler's third law, $P^2 \, \propto \, R^3$, in Eq. \eqref{sigmapiE_min} leads
    to the following scaling law:
    \begin{equation}
    \label{sigmapiE_min_R}
    \frac{\sigmapimin}{\pi_E} \left( \tEstar \right) \, \propto \, \beta^{1/2} \, R^{-1/4}.
    \end{equation}
    The maximum sensitivity to $\pi_E$ decreases slowly when one reduces the orbit
    radius, and what changes between a GSO and a LEO is just
    a factor of ${(6.6)}^{1/4} \approx 1.6$. Moreover, Eqs. \eqref{tEstar} and
    \eqref{tE} show that the timescale at which a LEO satellite is most sensitive
    to $\pi_E$ coincides with the typical Einstein timescale of an Earth-mass
    object.
    
    These properties are clearly shown in Fig. \ref{figure4}, where we consider
    a LEO satellite with HST-like orbital parameters. We assume an orbit radius
    $R = R_{\Earth} + 550 \, \text{km}$ ($P = 1 \, \text{h} \, 35.5 \, \text{min}$)
    and an event latitude $\lambda = 60^{\degree}$ for a bulge source,
    since the Hubble inclination above the equatorial plane is $28.5^{\degree}$.
    The predicted relative error $\sigmapi/\pi_E$ is plotted against the Einstein
    timescale $t_E$ for $D_L = 4 \, \text{kpc}$, $V = 200 \,
    \text{km/s}$, and zero blending ($\nu = 0$).
    We assume $f = 1 \, \text{observation}/ 3 \, \text{minutes}$ and
    $\sigma_m = 0.01$ as in Fig. \ref{figure1}, to allow for a quick comparison with
    the GSO. The plot shows the same physics as in Fig. \ref{figure1}, with
    two different regimes that arise for each choice of $\beta$. At $\beta = 0.1$,
    the maximum sensitivity to $\pi_E$ corresponds to Einstein timescales of
    one to several hours. According to Eq. \eqref{tE}, these $t_E$
    are typical of lens masses ranging from $M_{\Earth}$ to the
    super-Earth/ice giant transition. With a minimum
    relative error of roughly $10\%$, Fig. \ref{figure4} suggests that a detection
    of $\pi_E$ for these lenses should be possible for $\beta \lesssim 0.1-0.2$.
    Since finite source effects easily arise for these kinds of lenses
    (see Sect. \ref{finite_source})
    and can provide a measure of $\theta_E$, a LEO survey satellite could discover
    populations of free-floating objects ranging from terrestrial planets to
    super-Earths and ice giants. This would be a fantastic prospective for
    microlensing and exoplanet science in general. Clearly, a LEO survey would be affected by
    blending limitations, similar to the GSO. Moreover, the effects of Earth umbra
    have to be taken into account, because it reduces the fraction of the orbital period
    available to follow the source star (see Sect. \ref{earth_umbra}).
    
    For $\beta < 0.1$, Eqs. \eqref{tEstar} and \eqref{sigmapiE_min} are
    consistent with Fig. \ref{figure4}, similar to what is obtained for the
    GSO. In particular, we point out that with peak amplifications of several
    dozen, a LEO satellite should also be able to measure
    microlens parallax for $t_E \approx 1 \, \text{day}$ events, especially if
    ground-based observations are also available.

    \subsection{Finite source effects}
    \label{finite_source}
    Finite source effects arise when the impact parameter $\beta$ is comparable
    to $\rho = R_S x/R_E$, the projection of the source star radius $R_S$ onto
    the lens plane, measured in units of the Einstein radius $R_E$.
    Taking a Sun-like source as reference, the parameter $z \equiv \beta/\rho$ is
    given by
    \begin{equation}
    \label{z}
    z = 0.5
    \left( \frac{\beta}{0.1} \right)
    \left( \frac{t_E}{0.1 \, \text{days}} \right)
    \left( \frac{R_{\Sun}}{R_S} \right)
    \left( \frac{1/2}{x} \right)
    \left( \frac{V}{200 \, \text{km/s}} \right).
    \end{equation}
    Equation \eqref{z} shows that finite source effects are clearly measurable
    for $\beta = 0.1$ and $t_E = \tEstar \approx 0.1 \, \text{days}$. Moreover, 
    they are always detectable for $\beta < 0.1$ and $t_E = \tEstar$ since Eq.
    \eqref{tEstar} implies that $\beta \tEstar = \text{const}$.
    
    The Fisher analysis presented in Sect. \ref{Fisher_analysis} considers a point source.
    Can this assumption invalidate the error predictions of Fig. \ref{figure4}?
    The peak amplification influences $\sigmapi$ via the relation
    $A_{\text{max}} \simeq \beta^{-1}$. Therefore,
    finite source effects strongly modify the error predicted by a PSPL model
    if the maximum amplification of a finite source model,
    $A_{\text{max}}' = \sqrt{4 + \rho^2}/\rho$ \citep{Witt1994},
    is much smaller than $\beta^{-1}$, that is, $A_{\text{max}}'/A_{\text{max}}
    \ll 1$. Because $A_{\text{max}}'/A_{\text{max}} \simeq 2 \, z$, finite source effects
    strongly affect the error predictions only for $z \ll 1/2$ \citep{Witt1994}, which
    corresponds to $t_E \ll 0.1\, \text{days}$ at $\beta = 0.1$
    for source stars not much bigger
    than the Sun. As Fig. \ref{figure4} shows, these timescales are, however, of little
    importance since the error is already too big. For lower values of the impact
    parameter, $\beta < 0.1$, and $t_E \ll \tEstar$, the predictions are a
    priori strongly affected by finite source effects. However, since Fig. \ref{figure4}
    shows that for $t_E \ll \tEstar$, the sensitivity to $\pi_E$ rapidly saturates
    when one reduces $\beta$, we do not expect important deviations even for
    these timescales.
    
    In the case of giant-like source stars, $R_S \gtrsim 10
    \, R_\Sun$, finite source effects are dominant since peak amplifications can
    be much lower than what is predicted by a PSPL model. Basically, one can
    think in terms of an effective impact parameter $\beta' = \beta \, (2z)^{-1}$.
    For $R_S \gtrsim 10 \, R_\Sun$ and $t_E = 0.1\, \text{days}$, one has
    $A_{\text{max}}' \simeq 1$ and $\beta' \simeq 1$, and Fig. \ref{figure4}
    suggests that $\sigmapi$ would not guarantee a robust parallax detection. At
    $t_E = 1\, \text{day}$, one gets $\beta' \gtrsim 0.1$, and the parallax
    sensitivity is still marginal. Therefore, the finite source effects of giant-like
    stars seem to exclude robust measurements of $\pi_E$ for Earth-mass lenses.
    However, larger sources generally yield smaller photometric errors $\sigma_m$,
    and a parallax detection could still be feasible for closer lenses, i.e.,
    $D_L \lesssim 500 \, \text{pc}$ or $x \lesssim 0.06$, and $R_S \lesssim
    10 \, R_\Sun$ (see Eq. \ref{z}).

    \subsection{Earth umbra}
    \label{earth_umbra}
    The Earth shadow can represent a significant limitation to the capabilities
    of a LEO satellite. It reduces the fraction $\eta$ of the orbital period available
    to observe the microlensed source star. We calculate
    the dependence of this fraction on the orbital radius $R$ and the event latitude
    $\lambda$ over the orbital plane,
    \begin{equation}
    \label{umbra}
    \eta = 1 - \frac{1}{\pi} \arcsin \left[ \sqrt{1 - \left( \frac{R}{R_{\Earth}} \right)^2
    \sin^2 \lambda} \biggm/  \frac{R}{R_{\Earth}} \cos\lambda \right]
    \end{equation}
    For $R = R_{\Earth} + 550 \, \text{km}$, $\eta$ has a minimum value of $63\%$
    at $\lambda = 0^{\degree}$, and it slowly increases to $79\%$ for $\lambda =
    60^{\degree}$. For $\lambda \geq \lambda^{\star} \equiv \arcsin (R_{\Earth}/R)
    \approx 67^{\degree}$,
    the source star is always visible from the satellite. Consequently, even if the analysis
    of Sect. \ref{LEO} is strictly valid for continuous observations ($\eta = 1$), Eq.
    \eqref{umbra} seems to indicate that Earth umbra does not invalidate its general
    conclusions when one tracks bulge stars. There is also some room to minimize the
    impact of the Earth shadow by adjusting the satellite orbital parameters.
    For an altitude of 1000 km, for example, the angle $\lambda_{\star}$
    decreases to $60^{\degree}$. Moreover, increasing satellite inclination above the
    equatorial plane can substantially reduce the shadow impact.

\section{Conclusions}

   The present study employs a numerical Fisher matrix analysis to assess the
   feasibility of measuring microlens parallaxes by means of Earth orbit satellites.
   It extends the previous analytical analysis of \citet{Gould2013} to shorter Einstein
   timescales. We predict that, at $A_{\text{max}} \gtrsim 5-10$,
   a GSO satellite could detect microlens parallaxes for
   free-floating lenses with masses spanning from a fraction to a few dozen
   $\mjup$ (providing their mass if $\theta_E$ is also measured via finite
   source effects). It could
   also discover planets in tight orbits around very low-mass
   brown dwarfs. Moreover, at $\beta \lesssim 0.1-0.2$, a LEO satellite
   could discover free-floating objects ranging from terrestrial
   planets to super-Earths and ice giants. It could also detect, at $\beta \lesssim 0.05$,
   microlens parallaxes for Jupiter-mass free-floaters. Limitations to
   these results can be the strong requirements on the photometry ($0.01$
   magnitude error with three-minute exposures), the effects of blending,
   and in case of a LEO satellite, the Earth umbra. It is useful to note
   that, even though this study adopts a single-lens model, we extrapolated
   its results to binary-lens events. Indeed, as far as a parallax effect
   is concerned, the fundamental quantity is the projected separation
   between the source star and the lens centre of mass. In planetary
   events this almost coincides with the host. What really differentiates the two kinds
   of events is that finite source effects are needed to measure
   $\theta_E$ and estimate the lens mass: they are not routinely measured
   in single-lens events, while they are for binary ones.

\begin{acknowledgements}
      F. M. is grateful to N. Cornuault, L. Pittau, and C. Ranc for the frequent and
      fruitful discussions. The authors acknowledge the support of PERSU Sorbonne Université and the Programme National de Planétologie.
\end{acknowledgements}

\bibliographystyle{bibtex/aa.bst}    
\bibliography{bibtex/RotationMicrolens.bib}    

\end{document}